\documentclass[sigconf]{acmart}

\AtBeginDocument{%
  \providecommand\BibTeX{{%
    \normalfont B\kern-0.5em{\scshape i\kern-0.25em b}\kern-0.8em\TeX}}}

\setcopyright{acmcopyright}
\copyrightyear{2020}
\acmYear{2020}
\acmDOI{10.1145/1122445.1122456}

\acmConference[]{CoDS-COMAD'20}{January 05--07, 2020}{Hyderabad, India}
\acmBooktitle{7th ACM iKDD CoDS and 25th COMAD (CoDS-COMAD'20),
  January 05--07, 2020, Hyderabad, India}
\acmPrice{15.00}
\acmISBN{978-1-4503-9999-9/18/06}

\begin{document}

\title{Towards identifying Source credibility on Information Leakage in Digital Gadget Market}


\author{Kumari Neha}
\affiliation{%
  \institution{IIIT Delhi}
}
\email{nehak@iiitd.ac.in}

\author{Garvit Gupta}
\affiliation{%
  \institution{IIIT Delhi}
}
\email{garvit18141@iiitd.ac.in }

\author{Shreyans Mongia}
\affiliation{%
 \institution{IIIT Delhi}
}
\email{shreyans15178@iiitd.ac.in }

\author{Shubham Singh}
\affiliation{%
  \institution{IIIT Delhi}
}
\email{Shubham12101@iiitd.ac.in }

\author{Ponnurangam Kumaraguru}
\affiliation{%
  \institution{IIIT Delhi}
}
\email{pk@iiitd.ac.in}

\author{Arun Balaji Buduru}
\affiliation{%
  \institution{IIIT Delhi}
}
\email{arunb@iiitd.ac.in}


\renewcommand{\shortauthors}{Sample Text}

\begin{abstract}
The use of Social media to share content is on a constant rise. One
of the capsize effect of information sharing on Social media includes
the spread of sensitive information on the public domain. With the digital gadget market becoming highly competitive and ever-evolving, the trend of an increasing number of sensitive posts leaking information on devices in social media is observed. Many web-blogs on digital gadget market have mushroomed recently, making the problem of information leak all pervasive. Credible leaks on specifics of an upcoming device can cause a lot of financial damage to the respective organization. Hence, it is crucial to assess the credibility of the platforms that continuously post about a smartphone or digital gadget leaks. In this work, we analyze the headlines of leak web-blog posts and their corresponding official press-release. We first collect $54,495$ leak and press-release headlines for different smartphones. We train our custom NER model to capture the evolving smartphone names with an accuracy of $82.14\%$ on manually annotated results. We further propose a credibility score metric for the web-blog, based on the number of falsified and authentic smartphone leak posts.
\end{abstract}




\maketitle

\section{Introduction}\label{sec:intro1}
One of the capsize effect of information sharing on Social media is an abundance of alleged Information leaks of various sorts. The abundance of such alleged leaks can be credited to the growth in the number of people who post and consume news through social media. Among various social media platforms that post information leaks, one of the major sources are Web-blogs. In recent times, many web-blogs have mushroomed that feature posts about information leaks, especially related to smartphones. 
  We believe, the credible leaks on the upcoming devices can cause a lot of financial damage to the respective organization. Thus, it becomes crucial to continuously assess the credibility of the platforms that post about these leaks.

The study on information leak have previously focused on Organizational~\cite{10.1007/978-3-319-57454-7_14} as well as Personal Leaks~\cite{liu2010framework,yu2018leveraging}. The study of Organizational leaks has focused on assigning policy to documents which are shared within or outside the organization. While, the Users perspective on information leak focuses on privacy concerns for the users who share their information in the public domain. However, for our work, we define leak as any information of a smartphone or gadget which is not yet officially launched by the company, irrespective of the fact that the leak might be falsified later. The web-blogs have become a huge source of information about the ongoing projects as well as new releases for Smartphone companies. 
This is evident from the difference in Bounce rate\footnote{Bounce Rate is the percentage of single page visits to the website.} of different websites as shown in Table~\ref{table:alexa}, which shows the difference between the traffic received by the official press-release website as compared to the web-blogs which post leaks of the same product. 
We use Alexa~\footnote{\url{https://www.alexa.com/siteinfo/}}, to find out that the traffic on the web-blogs that post about leaked smartphone have greater percentage of visits of a single page view than the official press-release of the same smartphone. This indicates how pervasive the leak web-blogs have become in recent times.

Irrespective of authenticity, the leaks take away the opportunity of official unveiling of the product from the stakeholders~\cite{engadget}. This seals the fate of the smartphones before it had the chance to enter the market. One of the many effects these leaks have is, it can hamper the sale of products which are currently available in the market, while other being it gives away the competitors information of ongoing projects. The prevalence of these leaks are apparent; however, the authenticity of these leaks is still an open problem.

The headlines are the most crucial part of any web-blog post. They can not only attract more clicks; they can also reveal a lot about the blog post in a gist~\cite{Kourogi:2015:IAN:2806416.2806631,Reis2015BreakingTN,DBLP:conf/icwsm/PiotrkowiczDOM17}. We believe, since the leaks are supposed to attract more readers, they should have maximum information about the product in the headline itself. Therefore, for this study, we collect the headlines of the Web blogs that post about smartphone leaks and try to identify their credibility based on the official press-release information of the product. We consider the official press-release date-time as the actual release date of the smartphone. Any online appearance earlier than this date is considered as a leak.

\begin{table}[ht]
\centering
\begin{tabular}{|c|c|}
  \hline
  {\bf Website} &   {\bf Bounce rate}\\
  \hline
   notebookcheck.net & 71.8\\
  \hline
  androidpolice.com & 67.6 \\
  \hline
  \textbf{Samsung Press-releases} &56\\
  \hline
\end{tabular}
\caption{Table with the bounce rate(\%) of web-blogs which posts leak information about Samsung smartphone vs the official press-release page of Samsung}
\label{table:alexa}
\end{table}

With the goal to perform the credibility assessment of a given web-blog, we divide the problem at hand into following Research Questions:
\begin{itemize}
    \item \textbf{RQ1}: Does the headlines for leak posts differ from their press-release counter-part.
    \item \textbf{RQ2}: For a given product can we identify whether it was actually leaked, through the lense of our dataset.
    \item \textbf{RQ3}: For the given web-blog, can we continuously assess its credibility for information leaks.
\end{itemize}

The rest of the paper is structures as follows. We first look into the related work in Section~\ref{relatedWork}. Then we present the collected data and  preliminary analysis on the collected data in Section~\ref{dataDes}. In Section~\ref{proposedModel} we describe the proposed model. We finally discuss our results in the Section~\ref{results}. Lastly, we conclude our work with future work in Section~\ref{conclusion}.

\section{Related Work}\label{relatedWork}
The problem of Information credibility is a multi-facets problem. There is a fragile line between a rumour and information leak. The identification of the leaked content as sensitive or not adds another dimension to the problem at hand. Rumor can be defined as a statement whose truth value is unverifiable~\cite{Qazvinian:2011:RIM:2145432.2145602}. Rumors can further be classified as misinformation and disinformation~\cite{Fogg:1999:ECC:302979.303001}. The debunking of rumours and disinformation have become one of the most pervasive problems of the decade. While the early work of rumors and misinformation focuses on hand-crafted features~\cite{Qazvinian:2011:RIM:2145432.2145602,Castillo:2011:ICT:1963405.1963500,Liang2012ExpertFF,Liang2012ExpertFF}, the later works mainly focus on the deep learning approach to misinformation identification~\cite{kotonya-toni-2019-gradual,ma-etal-2018-rumor,Ma:2016:DRM:3061053.3061153}. The role of Twitter and other micro-blogging websites for misinformation spread has been widely studied in the literature~\cite{Castillo:2011:ICT:1963405.1963500,Liang2012ExpertFF}. Rumor detection approaches include content-based~\cite{Fogg:1999:ECC:302979.303001} and network based features~\cite{Fogg:1999:ECC:302979.303001,Castillo:2011:ICT:1963405.1963500}. Another approach also focuses on the expert knowledge in debunking of misinformation~\cite{Liang2012ExpertFF}. The study of rumors starts with some controversial topic~\cite{Fogg:1999:ECC:302979.303001} or trending topic~\cite{Castillo:2011:ICT:1963405.1963500}, where the role of human in the center can't be denied. 

One of the approaches to debunk the rumours is to identify the credibility of the source. One of the earliest definition of credibility describes it as a perceived quality composed of multiple dimensions~\cite{Fogg:1999:ECC:302979.303001}. While misinformation merely is false information, disinformation is deliberate false news~\cite{Fogg:1999:ECC:302979.303001}. The credibility assessment for topics in rumours have looked into factors including the reaction of specific topics, level of certainty of the news propagated, and whether URLs are included in the tweets~\cite{Castillo:2011:ICT:1963405.1963500}. 
In contrast to the tweets that have a limitation on the number of characters, the web-blogs and news posts don't pose any limitation to the word limit as per. While tweets have been extensively studied for rumour and stance identification, the news and blog articles for the study of rumours have been under-explored. News articles have been analyzed for effectiveness to attract clicks~\cite{Reis2015BreakingTN}. The headlines are related to the virality of the news article~\cite{Kourogi:2015:IAN:2806416.2806631}. The popularity of the news article can also be co-related to sentiment polarity~\cite{Reis2015BreakingTN}. We believe the headlines of a blog are supposed to give a gist of the content in the article. While the headlines are supposed to be informative, the perception of news can be changed with the help of the article~\cite{Horne2017TheIO}.

While the credibility of the information is of significant concern, it becomes a hotbed for the leak of sensitive information. The disclosures of sensitive information have been studied with respect to organization~\cite{10.1007/978-3-319-57454-7_14,Johnson2008InformationRO} as well as users~\cite{liu2010framework,Reis2015BreakingTN} in the past. The organizational leak of sensitive information deals with the sharing of documents from inside the organization to unintended people inside or outside the organization~\cite{10.1007/978-3-319-57454-7_14}. The authors in ~\cite{10.1007/978-3-319-57454-7_14} propose security policies for the sharing of the digital documents in the organization. They believe the policy will keep a check while the files are shared. The leak of sensitive personal information have been studied with respect to users posted information~\cite{liu2010framework} as well as Image sharing settings on Social Media~\cite{yu2018leveraging}. The work in ~\cite{liu2010framework} propose a framework to compute the privacy score of the user, which is indicative of the potential privacy risk due to participation in the network. 

Our work contributes to the given literature in two significant ways. To the best of our knowledge, we are first to investigate information leak of the product based companies and its credibility assessment. We believe our work is a small step towards the debunking of unauthorized as well as alleged information in the public domain.



\section{Problem Definition}\label{problemDef}
In our work, we deal with the credibility of the web-blogs who post about Information leaks about various smartphones. We believe the leak of credible information about a product before its official unveiling is a severe concern. Section~\ref{relatedWork} throws light on the importance of headlines on the spread and popularity of news. We believe the headlines act as an opening to the web-blog where the name of the product and its features are bound to be present. Thus, we consider the headlines of the web-blogs and press-releases for our analysis. 

\begin{figure}[!ht]
  \centering
  \includegraphics[width=\linewidth]{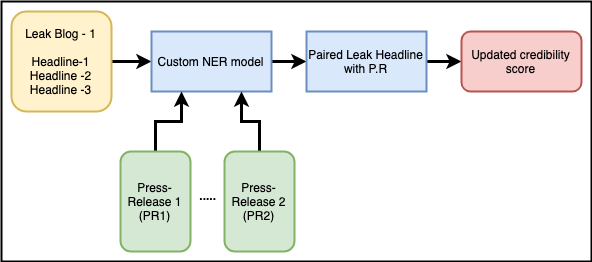}
  \caption{Architecture diagram of the proposed model. The first part consists of custom NER model. The second part groups product based on NER. In the third part we match the datetime of first appearance of leak and Press-release to assign the credibility score (P.R refers to Press-releases).}
  \Description{A}
  \label{fig:archi_infoleak}
\end{figure}

Assume we have a Web-blog that posts about smartphone leaks. Our objective can be divided into three parts:
\begin{enumerate}
    \item Extract smartphone names from the post headline with the help of our custom NER model. (\textbf{custom NER extraction})
    \item Group the smartphone leak headlines and press-release headlines into NER-based bins. (\textbf{NER-based binning}).
    \item Track the first online appearance through carbon-dating and produce smartphone's truth value(t)  for credibility calculation via Equation~\ref{credEquation}. (\textbf{Update Credibility Score})
    \begin{enumerate}
        \item Update the credibility score of the web-blog with every new smartphone leak post. 
    \end{enumerate}
\end{enumerate}
We present example of the headlines from web-blogs and press-releases in Table~\ref{table:sentencExp}. In the first objective, we deal with the evolving names of the smartphones in the market today. We train our NER extraction model on the top of Spacy's NER model~\cite{spacy2} in order to identify the smartphone names. We feed a list of smartphone names and custom sentence format to build a NER model that can generalize on smartphone-related headlines. 
The goal of the second task is to group the headlines according to predicted smartphone named-entities. The final task updates the credibility score of the web-blog with every smartphone leak falsified or verified. The final objective is to update  the credibility score of the web-blog dynamically with every product leak. The proposed architecture is shown in Figure ~\ref{fig:archi_infoleak}.

\section{Data Collection and Description}\label{dataDes}
The sample headlines of press-releases and leak blogs can be seen in Table~\ref{table:sentencExp}. We notice the press-release and leak blog headlines that talk about the same product show different semantic representation.
In this section, we first describe the collected data and further present some preliminary analysis on this data to understand the different semantic structures of the headlines.
\subsection{Data collection}
We first describe the collected data from different blogs as well as press-releases. We collected a total of $54,495$ headlines from the press-releases as well as the leak blog headlines. 
\subsubsection{Headlines of leak web-blogs}
We collect the headlines of $45,660$ posts from $6$ different blogging websites dedicated to  Smartphone leaks. We used Python's BeautifulSoup Library~\cite{b2s4Lib} to curate the headlines of leaks of a  wide range of smart devices. In Table~\ref{table:blogwiseLeak} we present the number of headlines collected from various web-blog sources with their average sentence length and vocabulary size. For the rest of the paper, we use the abbreviation for the web-blogs understudy as stated in the Table ~\ref{table:blogwiseLeak}
\begin{table*}[!ht]
\centering
\begin{tabular}{cccl}
  \toprule
  \textbf{Web-blog} &   \textbf{Number of Leak Headlines} & \textbf{Average Sentence Length(in Words)} & \textbf{Vocabulary Size}\\
  \midrule
  \textbf{AndroidPolice (AP)} & $1,428$ & $14.90$ & $2,873$\\
  \midrule
  \textbf{MySmartPrice (MSP)} & $634$ & $11.79$ & $1,222$\\
  \midrule
  \textbf{CultOfMac (COM)} & $2165$ & $10.83$ & $3,602$\\
  \midrule
  \textbf{MacRumors (MR)} & $39,921$ & $10.35$ & $18,701$ \\
  \midrule
  \textbf{wccftech (WCC)} & $1,000$ & $14.32$ & $2,441$\\
  \midrule
  \textbf{notebookcheck (NCR)} & $5,12$ & $12.24$ & $1,218$\\
  \bottomrule
\end{tabular}
\caption{Table with web-blog source, number of leak-headlines, average sentence length and vocabulary size in our dataset per blog.}
\label{table:blogwiseLeak}
\end{table*}

\begin{figure}[!ht]
  \centering
  \includegraphics[width=\linewidth]{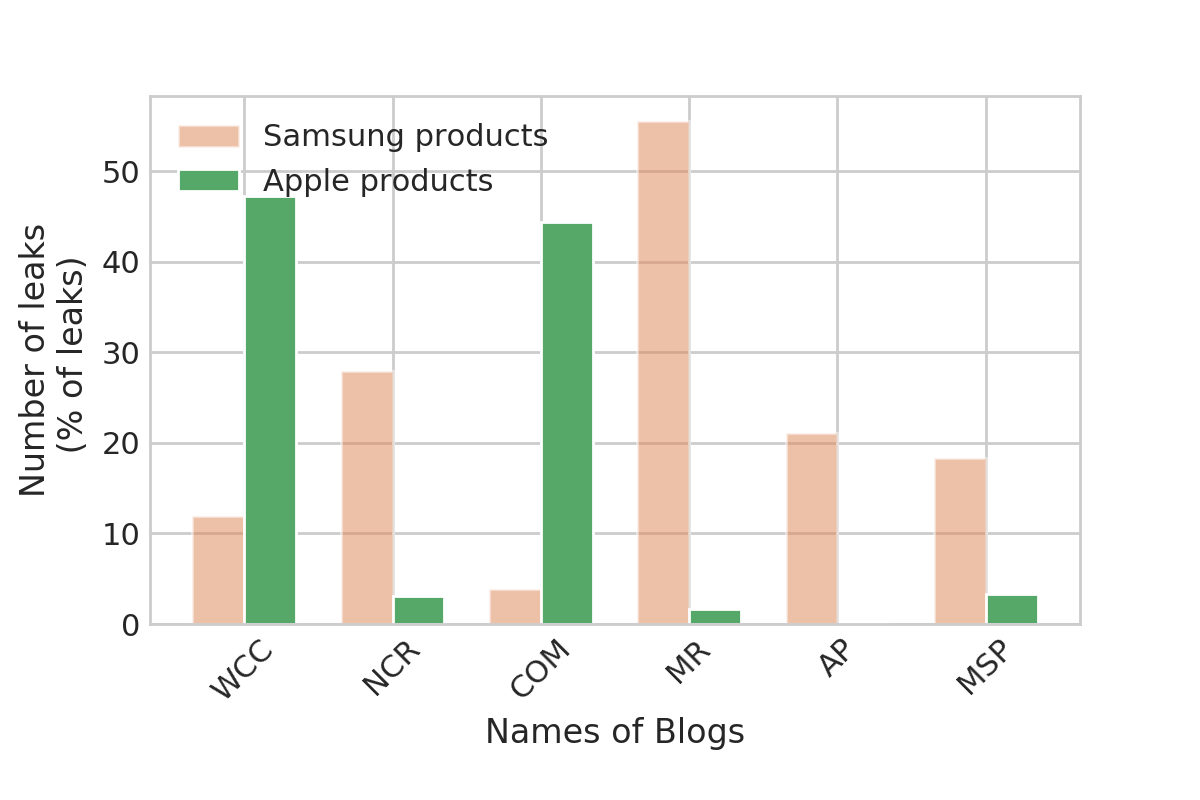}
  \caption{Bar chart of number of alleged smartphone leak by different web-blogs.
  }
  \Description{Bar chart of number of alleged smartphone leak by different web-blogs. We use abbreviations for the web-blogs as follows: AP is used for ANdroidpolice,COM is used for cultOfMac, MR is used for macRumors, wcc is used for wccftech and MSP is used for mySmartPrice.}
  \label{fig:samsung_apple_blogs}
\end{figure}

We plot the number of alleged leak smartphone headlines in different web-blogs under study for two smartphone companies in Figure \ref{fig:samsung_apple_blogs}.

\subsubsection{Headlines of official press-releases headlines}
In order to identify the truth behind the leaked blog headlines, we collect $8,835$ headlines from the official press-releases of  $14$ websites. The press-releases headlines talk about various smartphone launches as well as other company news.


\subsection{Preliminary Data Analysis:}
In this section, we try to understand the semantics of the press-releases and blog headlines and try to investigate their format.

\subsubsection{Sentiment Polarity}
\begin{figure}[!ht]
  \centering
  \includegraphics[width=\linewidth]{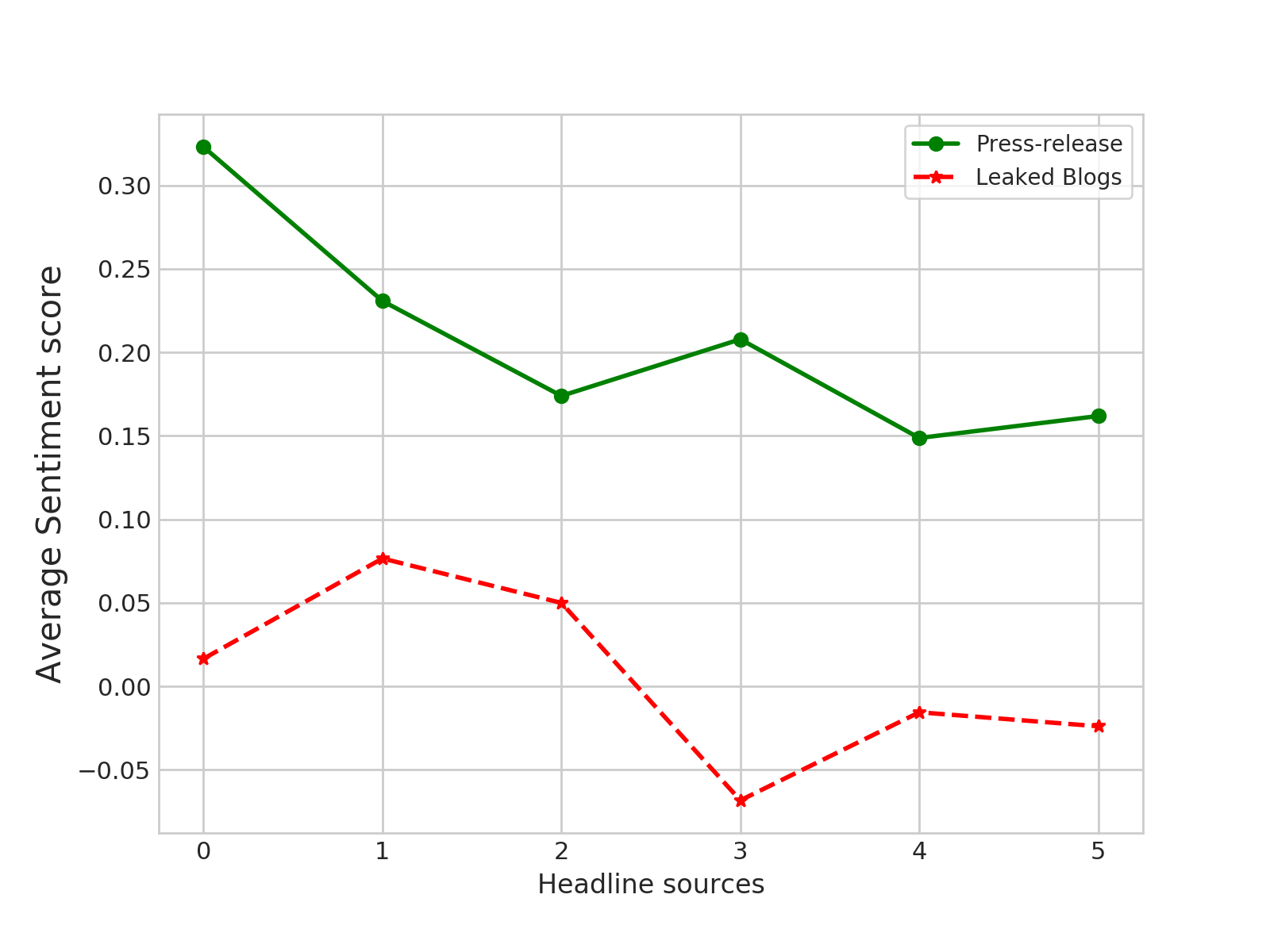}
  \caption{Comparison of the average compound sentiment score of leak headlines versus press-releases headlines}
  \Description{Comparison of the average compound sentiment score of leak blog headlines and press-releases headlines}
  \label{fig:senti}
\end{figure}

We use Vader~\cite{Hutto2014VADERAP} to understand how the sentiment polarity of the press-releases vary from the leak blog headlines.
The mean compound sentiment score for the leak blog headlines is $0.0025$ while for press-releases are $0.20$. This shows that press-releases of the product tends to post more positively about a product, unlike the leaked counterparts who tend to be less positive about it. We plot the average compound sentiment of the $6$ leak blog website as compared to 6 random press-release website's headlines in Figure ~\ref{fig:senti}. We clearly witness the average sentiment of blog and press-releases at opposite ends. 

\subsubsection{Active and Static Verb Analysis}
\begin{figure}[!ht]
  \centering
  \includegraphics[width=\linewidth]{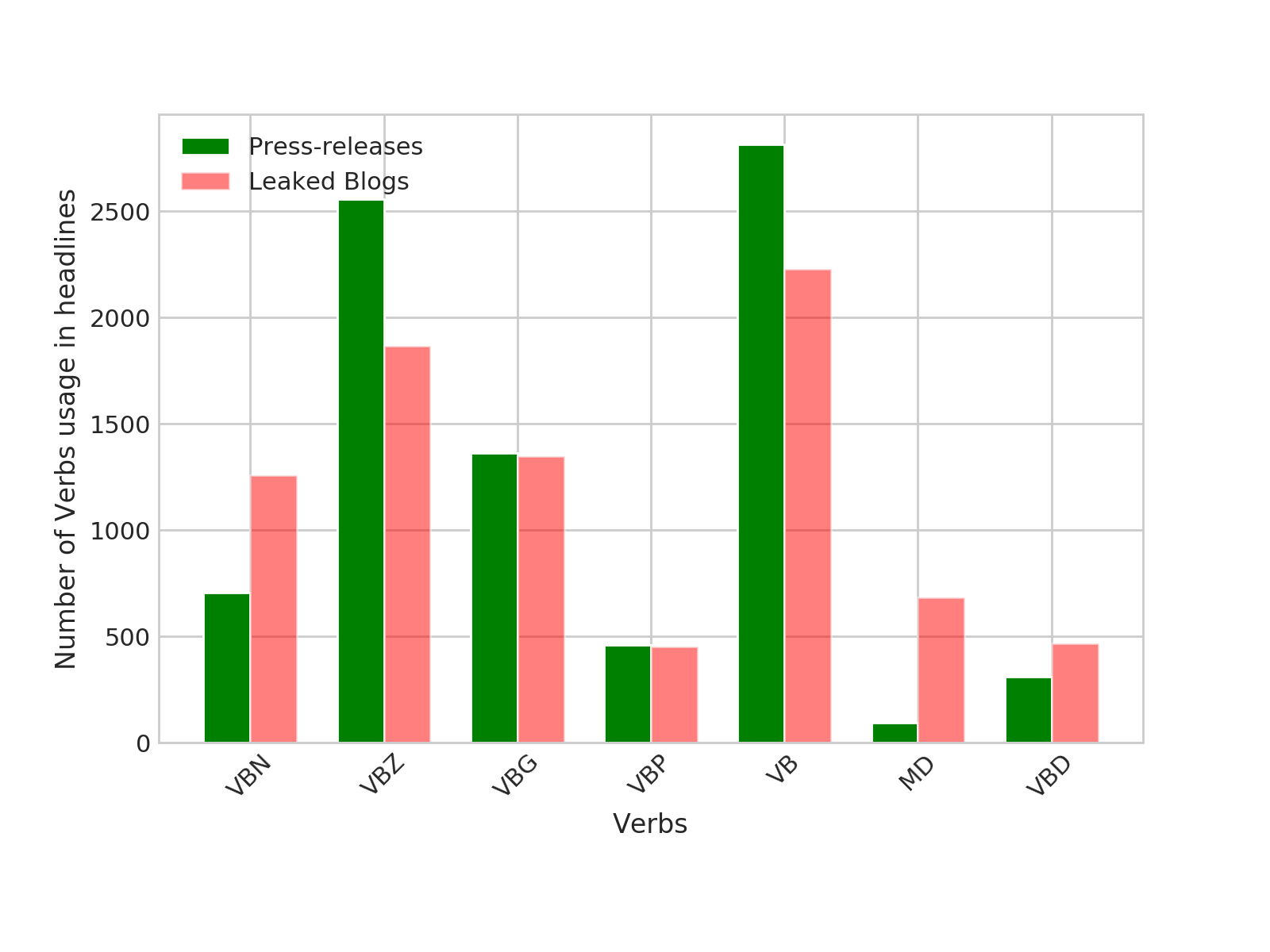}
  \caption{Distribution of verb form usage in press-releases and leaked blog headlines. Verb abbreviations: past participle as VBN, 3rd person singular present as VBZ, gerund or present participle as VBG, non-3rd person singular present as VBP, base form as VB, modal auxiliary as MD, past tense as VBD}
  \Description{Bar chart showing the Verb usage in press-releases vs leaked blog headlines. }
  \label{fig:verb}
\end{figure}

In order to analyze how the semantic structure of leak and press-releases headlines differ we plot the frequency of usage of different verb forms in Figure~\ref{fig:verb}. We witness that the use of past-participle as well as modal auxiliary verbs are more common in leak headlines, while the usage of base form of verb are more prominent in the press-releases. The use of non-3rd person singular present for both kind of sources however shows similar frequency of use. The use of past tense in verbs are also more common
in leak blogs than the press-releases. The usage of past Participle verbs shows the possibility that work is already complete. While the modal auxiliary verb indicates  the possibility of work going on. Thus, we believe, the leak web-blog structure their headlines mostly to talk about a developed product or the possibility of a new product in market.

\subsubsection{Lexical Structure of Headlines}
\begin{figure}[!ht]
  \centering
  \includegraphics[width=\linewidth]{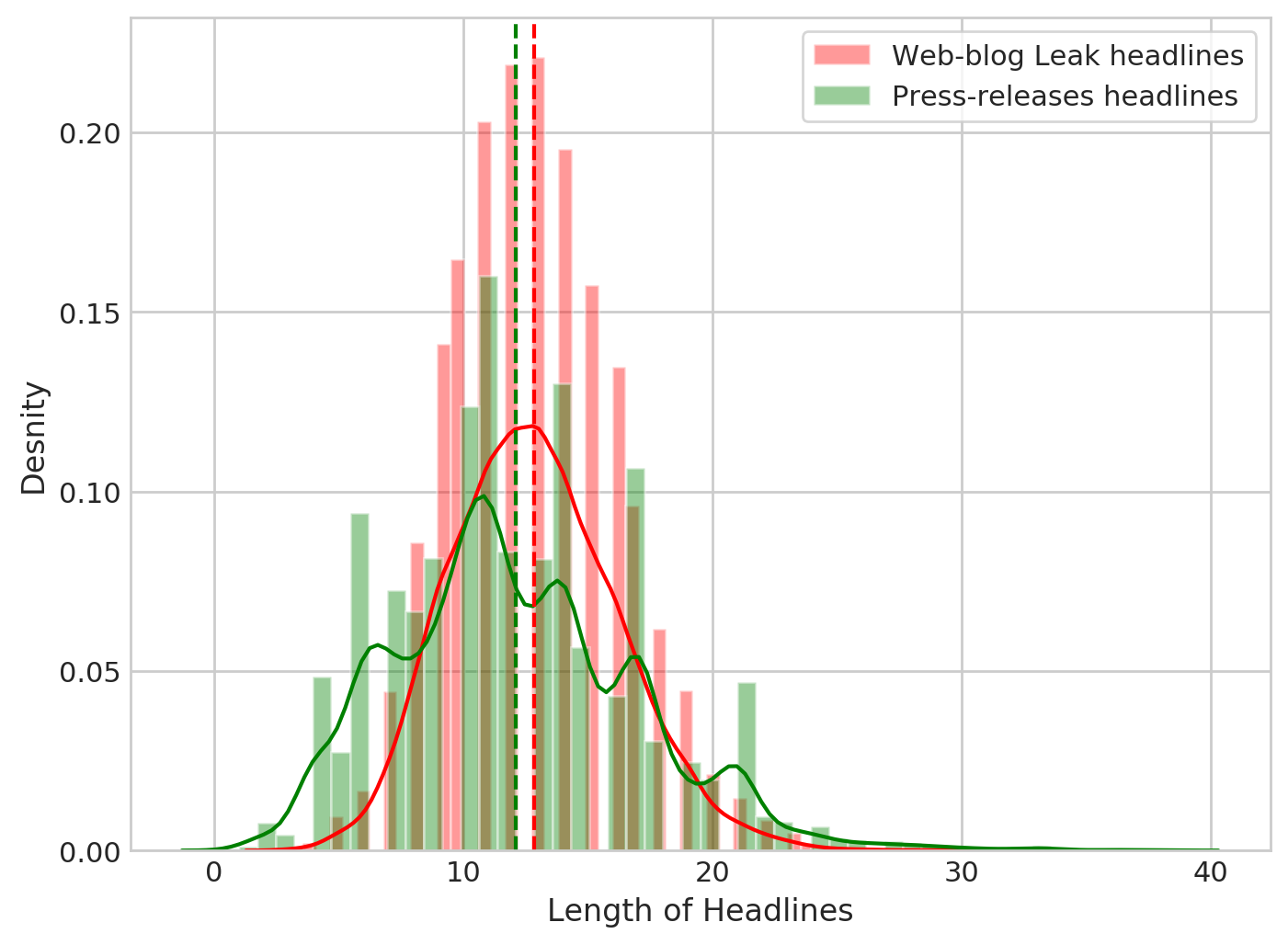}
  \caption{Distribution of the average length of the Press-release versus web-blog leak headlines. The mean length of both the headlines are approximately $12\%$. }
  \Description{Distribution of the Average length of the Press-release versus web-blog leak headlines. The mean length of both the headlines are approximately $12\%$.}
  \label{fig:length_headlines}
\end{figure}

We analyze the lexical structure of the leak web-blog headlines and press-release headlines to see if they share a common format or not. The mean length of the headlines for press-releases and web-blogs are $12.84\%$ and $12.10\%$ respectively.  The median value of the headlines for the leak web-blog is $13.0$ with a standard deviation of $3.34$. While the median of press-releases headlines is $11.0$ with standard deviation of $4.91$. The distribution of how the length of the two headlines varies is presented in Figure~\ref{fig:length_headlines}. The larger value of standard deviation for press-releases signifies that the press-releases headline posts have a bigger spread than the web-blog leaks. This shows that the press-releases of a smartphone company do not stick to a particular format while posting online; however the web-blogs have a comparatively limited format of making posts. 

\textbf{\textit{Custom stop words for smartphone headlines.}}
The headlines from both the press-releases and leak blog share a pattern in which certian words have similar likelihood of appearence. The example of such words are ``allege'', ``announce'', ``detail'', ``exclusive'',``leak'' etc. Since these words do not add up to the knowledge of smartphone feature or names, we added these words to the set of stopwords. We call these stopwords as custom stopwords. We manually curate a list of $251$ stopwords, of which the base form comes down to $220$. We use the custom stopwords on the top of the Spacy's Stopwords~\cite{spacy2} to remove the irrelevant keywords from the headlines when needed.

\begin{table}
\centering
\begin{tabular}{p{2cm}p{6cm}} 
\toprule
& {\bf Sentences} \\
\midrule
Leak & The 5G version of the Galaxy S10 will launch as a Verizon exclusive \\
Press-release & [Video] Hands-On with Galaxy S10 5G at MWC19\\ 
\midrule
Leak & Latest Samsung Galaxy S10 leak alleges inclusion of Infinity-O display\\
Press-release & [Interview] Infinity Viewing Experience: Behind the Galaxy S10 Display\\
\midrule
Leak & Leaked photos reveal new 2019 iPhone XR colors \\
Press-release &  Apple introduces iPhone XR \\
\bottomrule
\end{tabular}
\caption{Sample sentence pairs of leak and Press-release Data}\label{table:sentencExp}
\end{table}

\textbf{\textit{Carbon Dating Headline URLs.}}
In order to match the dates of the leak blogs and the press-releases, we need the date and time for each post. However, the date-time of when the web-blogs or press-releases are posted are either not available or are incomplete, where the time might be missing. In order to identify the credibility of a leak blogs, we consider the first online appearance of the leak blog headlines as an alleged leak. We use Carbon Dating~\cite{SalahEldeen:2013:CDW:2487788.2488121}, which polls the creation date from a number of sources and returns them in a machine-readable format. For most of the web-pages, there can be a considerable time difference between when a page was created, and when it was archived. Thus, the application built-in ~\cite{SalahEldeen:2013:CDW:2487788.2488121} is useful because the age estimation cannot rely simply on the web archives. Also, the web archive might have quarantined the release of specific holdings until some amount of time has passed. Thus the use of web-archive might not give the best results. 
The method that the Carbon Dating adopts is; first, they send a request to the host server and parse the output. Secondly, the authors use many backlinks to pull the timestamp of the URL. The timestamp includes Google search API and Social Media backlink. The first arrival time for the headline URLs are given by the below equation
\begin{equation}
    \label{carbonData}
    t_{estimated} = min(t_{method})
\end{equation}

Here, the $t_{estimated}$ is the first arrival time of the URL on the web and $t_{method}$ is the list of all the estimated times pulled out from various sources. 

\section{Approach}\label{proposedModel}
In order to identify the credibility of the web-blog source, we divide the proposed architecture into three parts, as discussed in Section~\ref{problemDef}. In this Section, we look at the proposed architecture in details. First, we look into the Custom NER extraction model for evolving smartphone names. 

\subsection{Custom NER model}
\begin{figure}[h!]
  \centering
  \includegraphics[width=\linewidth]{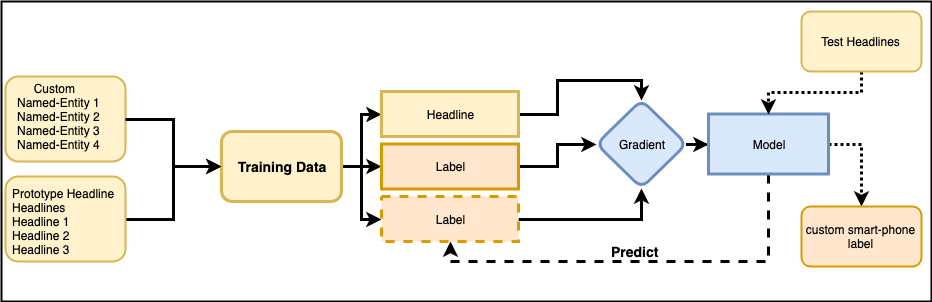}
  \caption{Diagram with the flow of custom NER model build on the top of Spacy~\cite{spacy2}. We provide the model with the sample training data along with the  smartphone labels.}
  \Description{A}
  \label{fig:ner_model}
\end{figure}

We witness the smartphone companies always come up with new product names. Since these smartphone names evolve with time, we found the existing NER model fails to identify all the smartphone names in the article's headline. Thus, we propose to update the present NER model that adapts to the custom Named-Entities based on the headline format of the leak blogs and press-releases headlines as shown in Figure ~\ref{fig:ner_model}. 

We use the available NER model from Spacy~\cite{spacy2} that provides an inbuilt statistical model for named-entity recognition. In order to train our own model on the top of Spacy, we provided the training data with Named-Entity in the format as shown in Table~\ref{table:example_sentence_train}.  Our training data consists of  $2,61,995$ headlines and $1,198$ smartphone names. We form the $2,61,995$ headlines from $84$ generic sentences for training our custom NER model.
The Spacy NER model has support for online training of the model. We start with a blank NER model and add entity recognizer in the pipeline. We add custom entity-labels to the entity recognizer. We create $17$ entity labels each corresponding to the name of the smartphone company. We run the model for $35$ iterations and use active learning to update the weights if the predicted value does not match with the annotated results. In order to avoid over-fitting, we use the drop-out rate of $0.35$, which have been the selected value for the drop-out rate for training NER model in Spacy. To prevent the model from generalization on the order of fed sentences, we shuffle the sentences in each of the $35$ iterations.
Our custom NER model took approximately $5.03$ days to train on a CUDA based implementation. We show the sample headline used for training our model in Table~\ref{table:example_sentence_train}. As shown in the Table~\ref{table:example_sentence_train}, the first column is the sentence from which we have to extract the named-entity, while the second column is the format in which the data is fed to the model.
\begin{table*}[!ht]
\centering
\begin{tabular}{|cl|} 
  \hline
  {\bf Sentence} &   {\bf NER}\\
  \hline
  \textbf{XXX release date, specs, features and rumours} & {`entities': [(0, n, `Samsung\_sp')]}\\
  \hline
  \textbf{[Update: More images] XXX leaked with slim bezels, no physical home button} & {`entities': [(n, n+t, `Apple\_sp')]}\\
  \hline
  \textbf{XXX leaks show ToF camera, 5x optical zoom, and multiple colors} & {`entities': [(0, n, `Samsung\_sp')]}\\
  \hline
\end{tabular}
\caption{Table with format of the sample training data fed to the custom NER model. Here XXX represents the name of the smartphone being talked about while `n' represents the location of the start of work and `t' denotes the displacement of the word.}
\label{table:example_sentence_train}
\end{table*}

\subsubsection{Efficiency of custom NER model}
In this section, we describe the efficiency of our trained NER model. 
To calculate the accuracy of the custom NER model, we use Jaccard similarity and Cosine similarity metrics and compare the result of the custom NER model with the annotated smartphone names.  We also compare the results of the similarity metrics to see which similarity metric gives the best result. We hypothesize the similarity metrics may perform differently for different leak blogs, given the way to form sentences may differ from blog to blog.
We use inter-annotation agreement to get the average value of similarity score for different similarity metric. The result of different similarity metrics are shown in Table~\ref{table:ner_similarity}. 

\textbf{Jaccard Similarity: }
We use Jaccard Similarity as a metric to find out the intersection versus union of the predicted NER and the annotated NER values. 
In order to find accurate result, we lemmatize the words before we calculate the Jaccard Score. The result of the Jaccard Similarity for the custom NER is given in Table~\ref{table:ner_similarity}.

\textbf{Cosine Similarity: }
We use pre-trained word vectors from Fasttext~\cite{mikolov2018advances} to vectorize the predicted and annotated NER for finding their cosine similarity. Mathematically, cosine similarity is the measures of cosine of the angle between two vectors which are projected in a multi-dimensional space.

\textbf{Annotation: }
We asked $2$ annotators to go over all the headlines in the dataset for the test set and mark the predicted NER as ``1'' if the name of smartphone appears in the predicted result and ``0'' otherwise. This annotation scheme is used to decide the accuracy of the predicted smartphone names by the model. 
The annotation scheme is also used to decide the Jaccard and Cosine similarity score for the grouping of leak headlines with the smartphone headlines, as mentioned in the second task in Section~\ref{problemDef}. In order to keep the prediction soft, we pre-process the predicted NER to remove the custom stopwords. 

\textbf{Inter-annotation agreement: } To calculate the inter-annotation agreement, we compared the annotation of the the two annotators for our test set and calculate the \textbf{$\kappa$} coefficient. Equation~\ref{kapa} shows formulation of \textbf{$\kappa$} statistics
\begin{equation}
    \label{kapa}
    \textbf{$\kappa$} = \frac{Pr(a) - Pr(e)}{1 - Pr(e)}
\end{equation}
where $Pr(a)$ denotes relative observed agreement between the two annotations, while $Pr(e)$ is the probability that the annotators agree by chance if they were randomly doing the annotation~\cite{krippendorff2004content}. Our annotation of set of $1,428$ test data shows $90.81\%$ agreement between the annotators.

\begin{figure}[!ht]
  \centering
  \includegraphics[width=\linewidth]{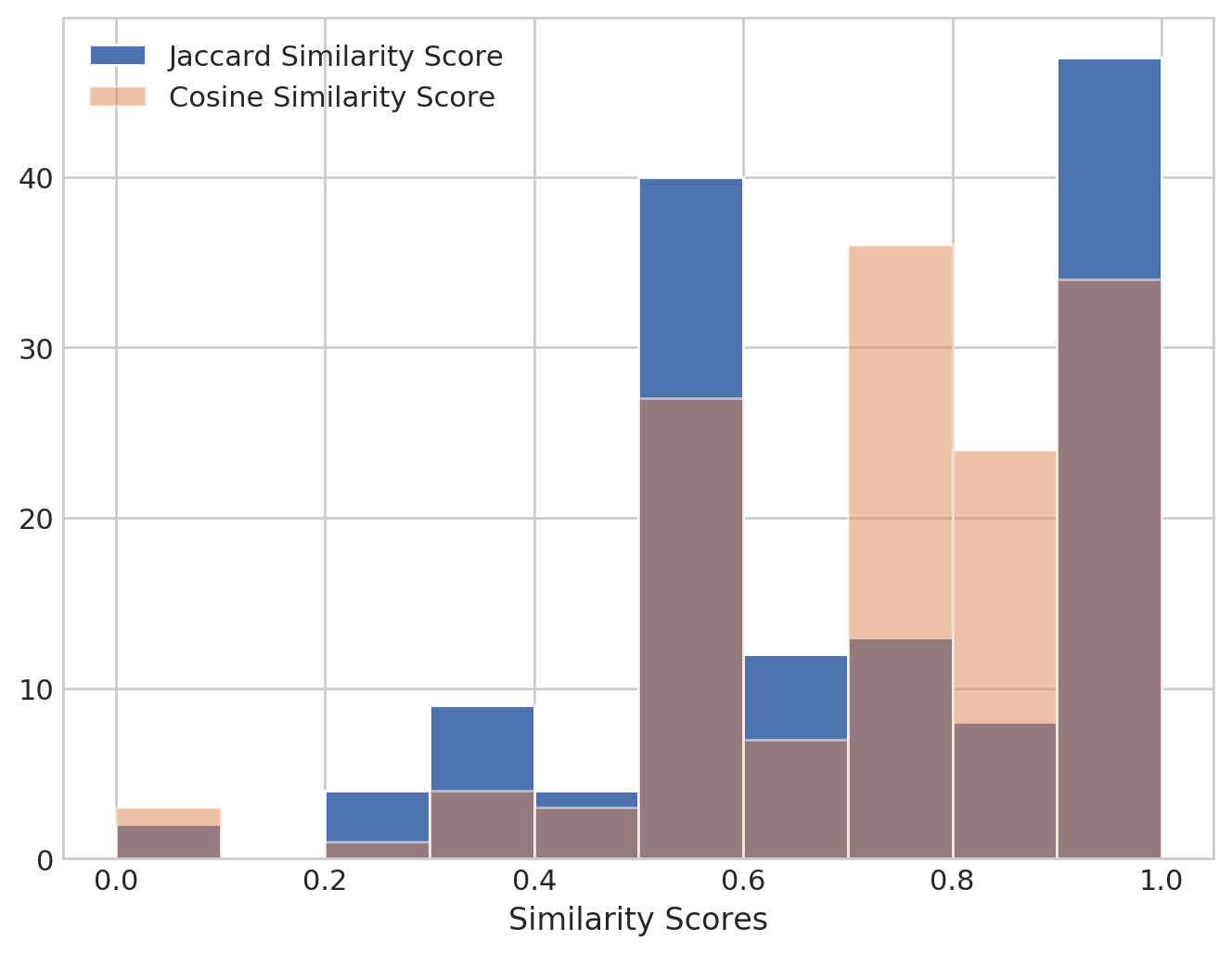}
  \caption{Comparison of the Jaccard similarity score and Cosine similarity score to visualize how the similarity score is distributed.}
  \Description{Comparison of the Jaccard similarity score and Cosine similarity score to visualize how the similarity score is distributed.}
  \label{fig:simcheck}
\end{figure}

Based on the headlines where the two annotators agree upon, we calculate the median of the similarity scores. The median of the Jaccard Similarity and Cosine Similarity came out to be 0.6 and 0.7, respectively. The distribution of the similarity metrics can be seen in the Figure~\ref{fig:simcheck}. We find the maximum score range from 0.2 to 1.0, where the 2 annotators agreed on the accuracy of the predicted model with the inter-annotation agreement of 90\%. However, given the normal distribution of the data for both leak and press-release headlines as shown in Figure~\ref{fig:length_headlines}, we looked into the number of right prediction on different percentile values. The goal here is to capture maximum named-entities in our dataset. We found the maximum number of accurate results were covered in the $25th$ percentile annotation. The $25th$ precentile value for both Cosine similarity and Jaccard similarity for our dataset is $0.5$. Therefore, we consider any NER prediction whose Jaccard similarity and Cosine similairty with the annotated named-entity came out to be above or equal to $0.5$, is true prediction.
%


\subsection{NER based binning}\label{ner_binning}
After we have identified the named-entity in the headlines, we group all the headlines based on the predicted smartphone names. Further, we use the datetime of release as a metric to identify whether the datetime of official press-release preceded alleged leak datetime or not. The $t$ and $f$ used in Equation~\ref{credEquation}, are thus updated on the basis of results. The $t$ and $f$  can have values $0$ or $1$, where the default values are taken as $0$. If the leaked headline's datetime precedes official press-release headline, the $t$ value is incremented or else $f$ value is incremented.

\subsection{Credibility Score for web-blogs}
After we have identified the truth behind the alleged leak, we update the score of the leak blog in our system. The default credibility score of any web-blog is taken as $0$. Any value greater that 0 will indicate towards source being authentic leak provider, while any score less would indicate the leak blog websites make inauthentic claims or rumors.
The goal here is to identify the credible sources who post information leak. The advantage of identifying the credibility of the web-blog source is two-fold. Firstly, further action can be taken on these web-blog sources by the respective companies in order to tackle the leaks of their smartphones. The second advantage is that the common consumers of the such news blogs can be benefited as they can rest assured the news of leak is not hoax.

We define the credibility score metric for a web-blog as given below:
\begin{equation}
    \label{credEquation}
    f = \frac{1}{n}\left(\sum_{k=1}^{n}{(t - 2*f)} \right)
\end{equation}
Here, n is the number of alleged leaked smartphones, $t$ is the binary value for the smartphone leak to be accurate based on the datetime as described in Section~\ref{ner_binning}, while f is the binary values for the given smartphone leak false or undefined in the press-releases. With every false leak claim, the credibility score of the blog decreases and with every valid claim, the credibility score of the web-blog increases. Overall, an adverse credibility score would mean that the credibility of the blog is low and most of the smartphone news published are rumours. While a positive credibility score will show, the leaks posted by web-blog are legit. We update the credibility score of the web-blog with every new article posted. This way, the credibility of the web-blog source can be kept in check.

\section{Experimental Results}\label{results}
In this section, we describe the results obtained from the proposed approach in details. For further analysis, we consider one of the web-blog source who actively post about smartphone leaks. Given that the official press-release of the products of all companies are hard to get, for our experiments, we consider the press-releases and corresponding leak of one of the smartphone companies. We compare the press-releases and leaks of Samsung for the credibility score of the Web-blog. The Experiments are divided into steps mentioned below: 
\begin{enumerate}
    \item We first evaluate the accuracy of the custom NER model with help of different similarity metrics for the given web-blog.
    \item We bin the press-release and leak blog headlines that talk about the same product using similarity metrics. 
    \item We identify the time-line of a given product release in the public domain for both press-release and leak and identify whether the first appearance is leak or not.
    \item We evaluate the percentage of leaks by a given leak blog website and calculate the credibility score for it.
\end{enumerate}

\begin{table}
\centering
\begin{tabular}{ccl} 
\toprule
\textbf{Web Blog} & \textbf{Jaccard Similaity} & \textbf{Cosine Similarity}\\
\midrule
AP & 85.71\%& 82.14\% \\
\bottomrule
\end{tabular}
\caption{Table showing the result of the similarity metric for NER model for \textbf{AP}.}
\label{table:ner_similarity}
\end{table}





\begin{figure}[!ht]
  \centering
  \includegraphics[width=\linewidth]{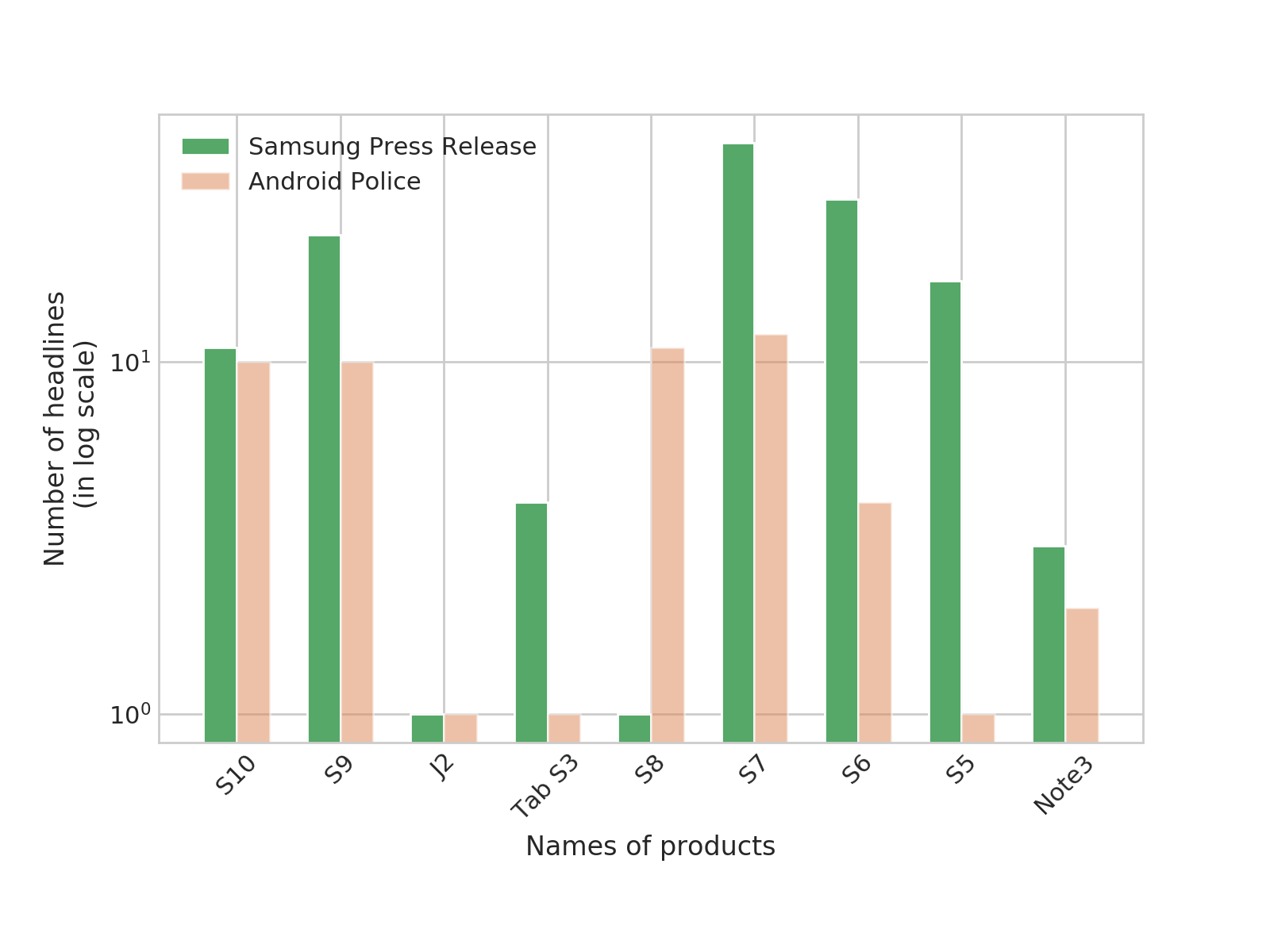}
  \caption{Distribution of the number of leaks and press-releases for different smartphones in the web-blog \textbf{AP}}
  \Description{Distribution of the number of leaks and press-releases for different smartphones in the web-blog \textbf{AP}}
  \label{fig:comparative_study_of_samsung}
\end{figure}

We use \textbf{AP} as our test case for further analysis. The further procedure are described in details as given below:

\textbf{Step I}: Identify number of smartphones talked about based on custom NER:

In order to identify the credibility of \textbf{AP}, we first analyze the number of product-wise leak of different smartphones in the web-blog as compared to the press-releases. We run our custom NER model to group all the headlines in \textbf{AP} according to predicted smartphone names. We then map the group with their respective press-release headline. We plot the number of headlines for different Samsung smartphones in the blog \textbf{AP} as well as in press-releases in the Figure~\ref{fig:comparative_study_of_samsung}. In our collected dataset, we witness the number of press-releases posts for all the smartphones is much more than the posts made by web-blog \textbf{AP}. While $S7$ is the most posted about product in both press-releases and web-blogs the number of posts for leak and press-releases differ massively for $S5$.

\textbf{Step II}: Group the alleged leak headlines with P.R:
After we have the identified the smartphone names talked about in \textbf{AP}, we bin all the headlines for a given smartphone together w.r.t both \textbf{AP} and Samsung press-releases. We finally sort the headlines based on the date and time obtained from the carbon-dating of URLs. We present press-release as well as the first online appearance of sample smartphones in the Table~\ref{AP_final}
\begin{table}
\centering
\begin{tabular}{p{1cm}p{3.0cm}p{3.0cm}}
\toprule
\textbf{Product Name} & \textbf{First Leak Datetime} & \textbf{First Press-release Datetime}\\
\midrule
S10 & July,$07$ $2018$ $07:54:44$ & Feb,$20$ $2019$ $08:59:56$ \\
\midrule
S9 & Dec,$15$ $2017$ $04:55:38$ & Feb,$25$ $2018$ $15:29:23$ \\
\midrule
S8 & Jan,$31$ $2017$ $09:22:28$ & Feb,$25$ $2017$ $13:54:13$ \\
\bottomrule
\end{tabular}
\caption{Table showing first appearance datetime(UTC) of the alleged leak and Press-release of smartphones for \textbf{AP}.}
\label{table:leakornot}
\label{AP_final}
\end{table}

\textbf{Step III}: Create timeline for product and give credibility score:

In the final step, we update the credibility score of the web-blog based on the number of leaks they have posted. More the negative value of the blog, less credible the source is.
The web-blog \textbf{AP}, posted about $15$ smartphones from Samsung. We calculate the credibility score of \textbf{AP} based on the n value as $15$ in Equation~\ref{credEquation}. The credibility score of \textbf{AP} came about to be $0.5$. Since the credibility score came out to be a positive value, we can say that the leak blog source is credible. However, since the credibility score calculation is a continuous process, it may vary as new leak posts are added to the leak blog website. 

\section{Conclusion}\label{conclusion}
In this work, we address the problem of identification of the credibility of the web-blog source who post about the leak of the smartphones.
We consider the leak of a smartphone as the post of a smartphone in the public domain before its official press-release. Our contributions in this paper are three-fold: $(1)$ We argue that the semantics of leak headlines are inherently different from the press-release headlines. $(2)$ We identify the truth behind the alleged leak with the help of the first appearance of the URLs of the headlines on the web. $(3)$ Based on the number of actual leaks by a web-log, we give the web-blog credibility score, which helps to identify whether the web-blog share legitimate leak information or not. 
We first collect the headlines of the web-blogs that talk about smartphone leaks. We then collect the press-releases of the smartphones from the official archive page. In order to identify the ever-evolving smartphone names, we train our custom NER model on the top of the Spacy. We manually annotate the headlines of the web-blog and present the accuracy of the custom NER model through various similarity metrics. We achieve the accuracy of $82.14\%$ on our custom NER model. In the next step, we group the headlines based on the NER and identify the created date and time of the leaked web-blog are before or after the press-release to identify the legitimate leaks. We propose a credibility score metric based on the number of smartphones the web-blog talk about and the number of identified leaks to give credibility score to the web-blog. 

We believe the credibility assessment of the web-blog is useful in two significant ways. Firstly, it can help the smartphone producing companies to identify the web-blogs who claim to have posted leaks. Secondly, it can help the users to identify legitimate sources for new information on smartphone leaks. We believe the leak of a smartphone is a severe issue and steps need to be taken for the identification of the alleged leaks. As a future work, we wish to extend the experiment of credibility assessment on all the blog sources under study and produce a comparative analysis of the same. We further wish to  improve the accuracy of the NER model to produce better results. We also wish to deploy a live web-blog credibility assessment system for the work done in this paper. 

\bibliographystyle{ACM-Reference-Format}
\bibliography{sample-base}


\begin{thebibliography}{22}


\ifx \showCODEN    \undefined \def \showCODEN     #1{\unskip}     \fi
\ifx \showDOI      \undefined \def \showDOI       #1{#1}\fi
\ifx \showISBNx    \undefined \def \showISBNx     #1{\unskip}     \fi
\ifx \showISBNxiii \undefined \def \showISBNxiii  #1{\unskip}     \fi
\ifx \showISSN     \undefined \def \showISSN      #1{\unskip}     \fi
\ifx \showLCCN     \undefined \def \showLCCN      #1{\unskip}     \fi
\ifx \shownote     \undefined \def \shownote      #1{#1}          \fi
\ifx \showarticletitle \undefined \def \showarticletitle #1{#1}   \fi
\ifx \showURL      \undefined \def \showURL       {\relax}        \fi
\providecommand\bibfield[2]{#2}
\providecommand\bibinfo[2]{#2}
\providecommand\natexlab[1]{#1}
\providecommand\showeprint[2][]{arXiv:#2}

\bibitem[\protect\citeauthoryear{Brad~Molen}{Brad~Molen}{2014}]%
        {engadget}
\bibfield{author}{\bibinfo{person}{www.engadget.com Brad~Molen}.}
  \bibinfo{year}{2014}\natexlab{}.
\newblock \bibinfo{title}{Leaks, lies and the bottom line}.
\newblock
\newblock
\urldef\tempurl%
\url{https://www.engadget.com/2014/08/08/smartphone-leaks-rumors/?guccounter=1}
\showURL{%
\tempurl}


\bibitem[\protect\citeauthoryear{Castillo, Mendoza, and Poblete}{Castillo
  et~al\mbox{.}}{2011}]%
        {Castillo:2011:ICT:1963405.1963500}
\bibfield{author}{\bibinfo{person}{Carlos Castillo}, \bibinfo{person}{Marcelo
  Mendoza}, {and} \bibinfo{person}{Barbara Poblete}.}
  \bibinfo{year}{2011}\natexlab{}.
\newblock \showarticletitle{Information Credibility on Twitter}. In
  \bibinfo{booktitle}{\emph{Proceedings of the 20th International Conference on
  World Wide Web}} \emph{(\bibinfo{series}{WWW '11})}.
  \bibinfo{publisher}{ACM}, \bibinfo{address}{New York, NY, USA},
  \bibinfo{pages}{675--684}.
\newblock
\showISBNx{978-1-4503-0632-4}
\urldef\tempurl%
\url{https://doi.org/10.1145/1963405.1963500}
\showDOI{\tempurl}


\bibitem[\protect\citeauthoryear{dos Reis, Benevenuto, de~Melo, Prates, Kwak,
  and An}{dos Reis et~al\mbox{.}}{2015}]%
        {Reis2015BreakingTN}
\bibfield{author}{\bibinfo{person}{J{\'u}lio~Cesar dos Reis},
  \bibinfo{person}{Fabr{\'i}cio Benevenuto}, \bibinfo{person}{Pedro O. S.~Vaz
  de Melo}, \bibinfo{person}{Raquel~Oliveira Prates}, \bibinfo{person}{Haewoon
  Kwak}, {and} \bibinfo{person}{Jisun An}.} \bibinfo{year}{2015}\natexlab{}.
\newblock \showarticletitle{Breaking the News: First Impressions Matter on
  Online News}.
\newblock \bibinfo{journal}{\emph{ArXiv}}  \bibinfo{volume}{abs/1503.07921}
  (\bibinfo{year}{2015}).
\newblock


\bibitem[\protect\citeauthoryear{Fogg and Tseng}{Fogg and Tseng}{1999}]%
        {Fogg:1999:ECC:302979.303001}
\bibfield{author}{\bibinfo{person}{B.~J. Fogg} {and} \bibinfo{person}{Hsiang
  Tseng}.} \bibinfo{year}{1999}\natexlab{}.
\newblock \showarticletitle{The Elements of Computer Credibility}. In
  \bibinfo{booktitle}{\emph{Proceedings of the SIGCHI Conference on Human
  Factors in Computing Systems}} \emph{(\bibinfo{series}{CHI '99})}.
  \bibinfo{publisher}{ACM}, \bibinfo{address}{New York, NY, USA},
  \bibinfo{pages}{80--87}.
\newblock
\showISBNx{0-201-48559-1}
\urldef\tempurl%
\url{https://doi.org/10.1145/302979.303001}
\showDOI{\tempurl}


\bibitem[\protect\citeauthoryear{Goyal, Mehta, and Srinivasan}{Goyal
  et~al\mbox{.}}{2017}]%
        {10.1007/978-3-319-57454-7_14}
\bibfield{author}{\bibinfo{person}{Tanya Goyal}, \bibinfo{person}{Sanket
  Mehta}, {and} \bibinfo{person}{Balaji~Vasan Srinivasan}.}
  \bibinfo{year}{2017}\natexlab{}.
\newblock \showarticletitle{Preventing Inadvertent Information Disclosures via
  Automatic Security Policies}. In \bibinfo{booktitle}{\emph{Advances in
  Knowledge Discovery and Data Mining}},
  \bibfield{editor}{\bibinfo{person}{Jinho Kim}, \bibinfo{person}{Kyuseok
  Shim}, \bibinfo{person}{Longbing Cao}, \bibinfo{person}{Jae-Gil Lee},
  \bibinfo{person}{Xuemin Lin}, {and} \bibinfo{person}{Yang-Sae Moon}} (Eds.).
  \bibinfo{publisher}{Springer International Publishing},
  \bibinfo{address}{Cham}, \bibinfo{pages}{173--185}.
\newblock
\showISBNx{978-3-319-57454-7}


\bibitem[\protect\citeauthoryear{Honnibal and Montani}{Honnibal and
  Montani}{2017}]%
        {spacy2}
\bibfield{author}{\bibinfo{person}{Matthew Honnibal} {and}
  \bibinfo{person}{Ines Montani}.} \bibinfo{year}{2017}\natexlab{}.
\newblock \bibinfo{title}{{spaCy 2}: Natural language understanding with
  {B}loom embeddings, convolutional neural networks and incremental parsing}.
  (\bibinfo{year}{2017}).
\newblock
\newblock
\shownote{To appear.}


\bibitem[\protect\citeauthoryear{Horne and Adali}{Horne and Adali}{2017}]%
        {Horne2017TheIO}
\bibfield{author}{\bibinfo{person}{Benjamin~D. Horne} {and}
  \bibinfo{person}{Sibel Adali}.} \bibinfo{year}{2017}\natexlab{}.
\newblock \showarticletitle{The Impact of Crowds on News Engagement: A Reddit
  Case Study}.
\newblock \bibinfo{journal}{\emph{ArXiv}}  \bibinfo{volume}{abs/1703.10570}
  (\bibinfo{year}{2017}).
\newblock


\bibitem[\protect\citeauthoryear{Hutto and Gilbert}{Hutto and Gilbert}{2014}]%
        {Hutto2014VADERAP}
\bibfield{author}{\bibinfo{person}{Clayton~J. Hutto} {and}
  \bibinfo{person}{Eric Gilbert}.} \bibinfo{year}{2014}\natexlab{}.
\newblock \bibinfo{title}{VADER: A Parsimonious Rule-Based Model for Sentiment
  Analysis of Social Media Text}.
\newblock
\newblock


\bibitem[\protect\citeauthoryear{Johnson}{Johnson}{2008}]%
        {Johnson2008InformationRO}
\bibfield{author}{\bibinfo{person}{M.~Eric Johnson}.}
  \bibinfo{year}{2008}\natexlab{}.
\newblock \showarticletitle{Information Risk of Inadvertent Disclosure: An
  Analysis of File-Sharing Risk in the Financial Supply Chain}.
\newblock \bibinfo{journal}{\emph{J. of Management Information Systems}}
  \bibinfo{volume}{25} (\bibinfo{year}{2008}), \bibinfo{pages}{97--124}.
\newblock


\bibitem[\protect\citeauthoryear{Kotonya and Toni}{Kotonya and Toni}{2019}]%
        {kotonya-toni-2019-gradual}
\bibfield{author}{\bibinfo{person}{Neema Kotonya} {and}
  \bibinfo{person}{Francesca Toni}.} \bibinfo{year}{2019}\natexlab{}.
\newblock \showarticletitle{Gradual Argumentation Evaluation for Stance
  Aggregation in Automated Fake News Detection}. In
  \bibinfo{booktitle}{\emph{Proceedings of the 6th Workshop on Argument
  Mining}}. \bibinfo{publisher}{Association for Computational Linguistics},
  \bibinfo{address}{Florence, Italy}, \bibinfo{pages}{156--166}.
\newblock
\urldef\tempurl%
\url{https://www.aclweb.org/anthology/W19-4518}
\showURL{%
\tempurl}


\bibitem[\protect\citeauthoryear{Kourogi, Fujishiro, Kimura, and
  Nishikawa}{Kourogi et~al\mbox{.}}{2015}]%
        {Kourogi:2015:IAN:2806416.2806631}
\bibfield{author}{\bibinfo{person}{Sawa Kourogi}, \bibinfo{person}{Hiroyuki
  Fujishiro}, \bibinfo{person}{Akisato Kimura}, {and} \bibinfo{person}{Hitoshi
  Nishikawa}.} \bibinfo{year}{2015}\natexlab{}.
\newblock \showarticletitle{Identifying Attractive News Headlines for Social
  Media}. In \bibinfo{booktitle}{\emph{Proceedings of the 24th ACM
  International on Conference on Information and Knowledge Management}}
  \emph{(\bibinfo{series}{CIKM '15})}. \bibinfo{publisher}{ACM},
  \bibinfo{address}{New York, NY, USA}, \bibinfo{pages}{1859--1862}.
\newblock
\showISBNx{978-1-4503-3794-6}
\urldef\tempurl%
\url{https://doi.org/10.1145/2806416.2806631}
\showDOI{\tempurl}


\bibitem[\protect\citeauthoryear{Krippendorff}{Krippendorff}{2004}]%
        {krippendorff2004content}
\bibfield{author}{\bibinfo{person}{K. Krippendorff}.}
  \bibinfo{year}{2004}\natexlab{}.
\newblock \bibinfo{booktitle}{\emph{Content Analysis: An Introduction to Its
  Methodology}}.
\newblock \bibinfo{publisher}{Sage}, \bibinfo{address}{USA}.
\newblock
\showISBNx{9780761915454}
\showLCCN{2003014200}
\urldef\tempurl%
\url{https://books.google.co.in/books?id=q657o3M3C8cC}
\showURL{%
\tempurl}


\bibitem[\protect\citeauthoryear{Liang, Liu, and Sun}{Liang
  et~al\mbox{.}}{2012}]%
        {Liang2012ExpertFF}
\bibfield{author}{\bibinfo{person}{Chen Liang}, \bibinfo{person}{Zhiyuan Liu},
  {and} \bibinfo{person}{Maosong Sun}.} \bibinfo{year}{2012}\natexlab{}.
\newblock \showarticletitle{Expert Finding for Microblog Misinformation
  Identification}. In \bibinfo{booktitle}{\emph{COLING}}.
\newblock


\bibitem[\protect\citeauthoryear{Liu and Terzi}{Liu and Terzi}{2010}]%
        {liu2010framework}
\bibfield{author}{\bibinfo{person}{Kun Liu} {and} \bibinfo{person}{Evimaria
  Terzi}.} \bibinfo{year}{2010}\natexlab{}.
\newblock \showarticletitle{A framework for computing the privacy scores of
  users in online social networks}.
\newblock \bibinfo{journal}{\emph{ACM Transactions on Knowledge Discovery from
  Data (TKDD)}} \bibinfo{volume}{5}, \bibinfo{number}{1}
  (\bibinfo{year}{2010}), \bibinfo{pages}{6}.
\newblock


\bibitem[\protect\citeauthoryear{Ma, Gao, Mitra, Kwon, Jansen, Wong, and
  Cha}{Ma et~al\mbox{.}}{2016}]%
        {Ma:2016:DRM:3061053.3061153}
\bibfield{author}{\bibinfo{person}{Jing Ma}, \bibinfo{person}{Wei Gao},
  \bibinfo{person}{Prasenjit Mitra}, \bibinfo{person}{Sejeong Kwon},
  \bibinfo{person}{Bernard~J. Jansen}, \bibinfo{person}{Kam-Fai Wong}, {and}
  \bibinfo{person}{Meeyoung Cha}.} \bibinfo{year}{2016}\natexlab{}.
\newblock \showarticletitle{Detecting Rumors from Microblogs with Recurrent
  Neural Networks}. In \bibinfo{booktitle}{\emph{Proceedings of the
  Twenty-Fifth International Joint Conference on Artificial Intelligence}}
  \emph{(\bibinfo{series}{IJCAI'16})}. \bibinfo{publisher}{AAAI Press},
  \bibinfo{pages}{3818--3824}.
\newblock
\showISBNx{978-1-57735-770-4}
\urldef\tempurl%
\url{http://dl.acm.org/citation.cfm?id=3061053.3061153}
\showURL{%
\tempurl}


\bibitem[\protect\citeauthoryear{Ma, Gao, and Wong}{Ma et~al\mbox{.}}{2018}]%
        {ma-etal-2018-rumor}
\bibfield{author}{\bibinfo{person}{Jing Ma}, \bibinfo{person}{Wei Gao}, {and}
  \bibinfo{person}{Kam-Fai Wong}.} \bibinfo{year}{2018}\natexlab{}.
\newblock \showarticletitle{Rumor Detection on Twitter with Tree-structured
  Recursive Neural Networks}. In \bibinfo{booktitle}{\emph{Proceedings of the
  56th Annual Meeting of the Association for Computational Linguistics (Volume
  1: Long Papers)}}. \bibinfo{publisher}{Association for Computational
  Linguistics}, \bibinfo{address}{Melbourne, Australia},
  \bibinfo{pages}{1980--1989}.
\newblock
\urldef\tempurl%
\url{https://doi.org/10.18653/v1/P18-1184}
\showDOI{\tempurl}


\bibitem[\protect\citeauthoryear{Mikolov, Grave, Bojanowski, Puhrsch, and
  Joulin}{Mikolov et~al\mbox{.}}{2018}]%
        {mikolov2018advances}
\bibfield{author}{\bibinfo{person}{Tomas Mikolov}, \bibinfo{person}{Edouard
  Grave}, \bibinfo{person}{Piotr Bojanowski}, \bibinfo{person}{Christian
  Puhrsch}, {and} \bibinfo{person}{Armand Joulin}.}
  \bibinfo{year}{2018}\natexlab{}.
\newblock \showarticletitle{Advances in Pre-Training Distributed Word
  Representations}. In \bibinfo{booktitle}{\emph{Proceedings of the
  International Conference on Language Resources and Evaluation (LREC 2018)}}.
\newblock


\bibitem[\protect\citeauthoryear{Piotrkowicz, Dimitrova, Otterbacher, and
  Markert}{Piotrkowicz et~al\mbox{.}}{2017}]%
        {DBLP:conf/icwsm/PiotrkowiczDOM17}
\bibfield{author}{\bibinfo{person}{Alicja Piotrkowicz}, \bibinfo{person}{Vania
  Dimitrova}, \bibinfo{person}{Jahna Otterbacher}, {and} \bibinfo{person}{Katja
  Markert}.} \bibinfo{year}{2017}\natexlab{}.
\newblock \showarticletitle{Headlines Matter: Using Headlines to Predict the
  Popularity of News Articles on Twitter and Facebook}. In
  \bibinfo{booktitle}{\emph{Proceedings of the Eleventh International
  Conference on Web and Social Media, {ICWSM} 2017, Montr{\'{e}}al,
  Qu{\'{e}}bec, Canada, May 15-18, 2017.}} \bibinfo{publisher}{{AAAI} Press},
  \bibinfo{address}{Canada}, \bibinfo{pages}{656--659}.
\newblock
\urldef\tempurl%
\url{https://aaai.org/ocs/index.php/ICWSM/ICWSM17/paper/view/15657}
\showURL{%
\tempurl}


\bibitem[\protect\citeauthoryear{Qazvinian, Rosengren, Radev, and
  Mei}{Qazvinian et~al\mbox{.}}{2011}]%
        {Qazvinian:2011:RIM:2145432.2145602}
\bibfield{author}{\bibinfo{person}{Vahed Qazvinian}, \bibinfo{person}{Emily
  Rosengren}, \bibinfo{person}{Dragomir~R. Radev}, {and}
  \bibinfo{person}{Qiaozhu Mei}.} \bibinfo{year}{2011}\natexlab{}.
\newblock \showarticletitle{Rumor Has It: Identifying Misinformation in
  Microblogs}. In \bibinfo{booktitle}{\emph{Proceedings of the Conference on
  Empirical Methods in Natural Language Processing}}
  \emph{(\bibinfo{series}{EMNLP '11})}. \bibinfo{publisher}{Association for
  Computational Linguistics}, \bibinfo{address}{Stroudsburg, PA, USA},
  \bibinfo{pages}{1589--1599}.
\newblock
\showISBNx{978-1-937284-11-4}
\urldef\tempurl%
\url{http://dl.acm.org/citation.cfm?id=2145432.2145602}
\showURL{%
\tempurl}


\bibitem[\protect\citeauthoryear{Richardson}{Richardson}{2004}]%
        {b2s4Lib}
\bibfield{author}{\bibinfo{person}{Leonard Richardson}.}
  \bibinfo{year}{2004}\natexlab{}.
\newblock \bibinfo{title}{{BeautifulSoup Library}}.
\newblock
  \bibinfo{howpublished}{\url{https://www.crummy.com/software/BeautifulSoup/bs4/doc/}}.
\newblock


\bibitem[\protect\citeauthoryear{SalahEldeen and Nelson}{SalahEldeen and
  Nelson}{2013}]%
        {SalahEldeen:2013:CDW:2487788.2488121}
\bibfield{author}{\bibinfo{person}{Hany~M. SalahEldeen} {and}
  \bibinfo{person}{Michael~L. Nelson}.} \bibinfo{year}{2013}\natexlab{}.
\newblock \showarticletitle{Carbon Dating the Web: Estimating the Age of Web
  Resources}. In \bibinfo{booktitle}{\emph{Proceedings of the 22Nd
  International Conference on World Wide Web}} \emph{(\bibinfo{series}{WWW '13
  Companion})}. \bibinfo{publisher}{ACM}, \bibinfo{address}{New York, NY, USA},
  \bibinfo{pages}{1075--1082}.
\newblock
\showISBNx{978-1-4503-2038-2}
\urldef\tempurl%
\url{https://doi.org/10.1145/2487788.2488121}
\showDOI{\tempurl}


\bibitem[\protect\citeauthoryear{Yu, Kuang, Zhang, Zhang, Lin, and Fan}{Yu
  et~al\mbox{.}}{2018}]%
        {yu2018leveraging}
\bibfield{author}{\bibinfo{person}{Jun Yu}, \bibinfo{person}{Zhenzhong Kuang},
  \bibinfo{person}{Baopeng Zhang}, \bibinfo{person}{Wei Zhang},
  \bibinfo{person}{Dan Lin}, {and} \bibinfo{person}{Jianping Fan}.}
  \bibinfo{year}{2018}\natexlab{}.
\newblock \showarticletitle{Leveraging content sensitiveness and user
  trustworthiness to recommend fine-grained privacy settings for social image
  sharing}.
\newblock \bibinfo{journal}{\emph{IEEE Transactions on Information Forensics
  and Security}} \bibinfo{volume}{13}, \bibinfo{number}{5}
  (\bibinfo{year}{2018}), \bibinfo{pages}{1317--1332}.
\newblock


\end{thebibliography}

\end{document}